\documentclass[twocolumn,showpacs,preprintnumbers,amsmath,amssymb]{revtex4}
\usepackage{graphicx}% Include figure files
\usepackage{dcolumn}% Align table columns on decimal point
\usepackage{bm}% bold math

\newcommand{\be}{\begin{equation}}
\newcommand{\ee}{\end{equation}}
\newcommand{\lb}{\label}
\newcommand{\Ol}{\overline}
\newcommand{\ba}{{\bf a}}
\newcommand{\bF}{{\bf f}}
\newcommand{\bn}{{\bf n}}
\newcommand{\br}{{\bf r}}
\newcommand{\bu}{{\bf u}}
\newcommand{\bx}{{\bf x}}
\newcommand{\bA}{{\bf A}}
\newcommand{\bS}{{\bf S}}

\newcommand{\bsigma}{{\mbox{\boldmath $\sigma$}}}
\newcommand{\btau}{{\mbox{\boldmath $\tau$}}}
\newcommand{\bomega}{{\mbox{\boldmath $\omega$}}}
\newcommand{\grad}{{\mbox{\boldmath $\nabla$}}}
\newcommand{\bdot}{{\mbox{\boldmath $\cdot$}}}
\newcommand{\btimes}{{\mbox{\boldmath $\times$}}}

\newcommand{\hn}{\hat{\bf n}}
\newcommand{\hT}{\hat{\bf t}}

\begin{document}

% \preprint{APS/123-QED}

\title{The Cascade of Circulations in Fluid Turbulence}

\author{Gregory L. Eyink}
\altaffiliation[Also at ]{Center for Nonlinear Studies, Los Alamos
 National Laboratory}
 \email{eyink@ams.jhu.edu}
\affiliation{%
Department of Applied Mathematics \& Statistics, The Johns Hopkins University\\
Baltimore, MD 21218
}%

\date{\today}% It is always \today, today,

\begin{abstract}
Kelvin's Theorem on conservation of circulations  is an essential ingredient of
G. I. Taylor's
theory of turbulent energy dissipation by the process of vortex-line
stretching. In previous
work, we have proposed a nonlinear mechanism for the breakdown of Kelvin's
Theorem
in ideal turbulence at infinite Reynolds number. We develop here a detailed
physical theory
of this ``cascade of circulations''.  Our analysis is based upon an effective
equation for
large-scale ``coarse-grained'' velocity, which contains a turbulent-induced
``vortex-force'' that
can violate Kelvin's Theorem. We show that singularities of sufficient
strength, which are observed
to exist in turbulent flow,  can lead to non-vanishing dissipation of
circulation for an arbitrarily
small filtering length in the effective equations. This result is an analogue
for circulation of
Onsager's theorem on energy dissipation for singular Euler solutions. The
physical mechanism of
the breakdown of Kelvin's Theorem is diffusion of lines of large-scale
vorticity out of the
advected loop. This phenomenon can be viewed as a classical analogue of the
Josephson-Anderson
phase-slip phenomenon in superfluids due to quantized vortex lines. We show
that the
circulation cascade is local in scale and use this locality to develop concrete
expressions
for the turbulent vortex-force by a multi-scale gradient-expansion.  We discuss
implications
for Taylor's theory of turbulent dissipation and we point out some related
cascade phenomena,
in particular for magnetic-flux in magnetohydrodynamic (MHD) turbulence.
% *Valid PACS numbers may be entered using the \verb+\pacs{#1}+ command.
\end{abstract}

\pacs{47.27.Ak,47.27.Jv,47.32.Cc,47.37.+q}% PACS, the Physics and Astronomy
\maketitle

\section{Introduction}
\lb{intro}

The fundamental laws of vortex motion for incompressible inviscid fluids in
three space dimensions were formulated by Helmholtz \cite{Helmholtz1858}.
Starting from the incompressible Euler equations for an ideal fluid, he showed
that  vortex lines are material lines and that the flux within any vortex tube
is a Lagrangian
invariant. Lord Kelvin \cite{Kelvin1869} gave an elegant alternative
formulation
of these laws in terms of the conservation of circulation, for any closed loop
advected
by an ideal fluid. This theorem is equally valid in any space dimension.

However, all of these results depend upon an implicit assumption that the
solutions of
the fluid equations remain smooth in the inviscid limit. In this limit, as the
Reynolds number
tends to infinity, all smooth, laminar solutions of the Euler equations are
unstable and
the fluid motion becomes turbulent. For infinite-Reynolds-number turbulent
solutions,
standard conservation laws of the ideal  Euler equations of motion need not
hold. For example,
both experiments \cite{Dryden43,Sreenivasan84, Cadotetal97, Pearsonetal02} and
simulations
\cite{Sreenivasan98,Kanedaetal03} show that energy is not conserved in
turbulent fluids
even in the limit as molecular viscosity tends to zero. The anomalous rate of
energy
dissipation in turbulent fluids was attributed by Onsager \cite{Onsager1949} to
predicted
H\"{o}lder singularities in the solutions of the inviscid Euler equations. In
particular, he showed
that a (spatially-minimum) H\"{o}lder exponent $h_{\min}\leq 1/3$ is necessary
for an
Euler solution to dissipate energy.  See also
\cite{Eyink1994,Constantinetal1994,
DuchonRobert2000, EyinkSreenivasan06}. The existence of such near-singularities
for turbulent velocity fields at high Reynolds number has been confirmed by
data from
experiments and simulations \cite{Muzyetal91,Arneodoetal95,KestenerArneodo04}.

In a previous work \cite{Eyink06} (hereafter referred to as ``I") we considered
similar
questions for the conservation of circulations by turbulent solutions. In that
paper
we proved an analogue of Onsager's theorem, stating necessary conditions for
the
anomalous dissipation of circulations by inviscid Euler solutions. Furthermore,
since
these conditions are expected to be satisfied in turbulent flow, we conjectured
that
Kelvin's Theorem, in its usual form, indeed breaks down for the relevant
high-Reynolds
number solutions. We termed this phenomenon a ``cascade of circulations." In a
following
paper \cite{Chenetal06} we presented evidence from direct numerical simulations
for
the existence of such a cascade. The purpose of the present paper is to
elaborate
further the physical theory of this phenomenon. In particular, our aims are as
follows:

In the remainder of this section of the paper, we shall discuss some important
background
information. We first remind the reader of the classical Kelvin Theorem. Next
we briefly
review some ideas of G. I. Taylor \cite{Taylor17,TaylorGreen37,Taylor1938}
about the
role of circulation-conservation in the production of energy dissipation in
three-dimensional
turbulence. In the second section of the paper we present our new results.
First, we discuss
the filtering approach which is the basis of our theory, and explain its
relation to
renormalization-group (RG) ideas and to large-eddy simulation (LES) modelling
of turbulent
flows. Second, we establish exact results for large-scale circulation balance
of low-pass
filtered velocity fields. Third, we explain how Taylor's argument can be
extended to stretching
of filtered vorticity and how this is related to forward cascade of energy
through the
inertial range. Fourth,  we review the results from I on  the possibility of
anomalous
dissipation of circulations in the limit of zero filtering length. Fifth, we
point out an
interesting analogy between this cascade of circulations and the phenomenon of
phase-slip in superfluids,  noting similarities with previous ideas of P. W.
Anderson
\cite{Anderson66}. Sixth, we discuss the scale-locality of the
circulation-cascade and
elaborate a multi-scale gradient (MSG) expansion for circulation-flux, along
the lines
laid out earlier for turbulent stress \cite{Eyink05,Eyink06a}. Finally, in the
conclusion
section we discuss some implications of our results and various extensions to
magnetohydrodynamic (MHD) and geophysical fluid turbulence.

\subsection{Classical Kelvin Theorem}
\lb{kelvin}

We here briefly review some standard facts about the conservation of
circulations.
Let $\bu(\bx,t)$ be a smooth velocity field solving the incompressible
Navier-Stokes
equation with viscosity $\nu$
\be \partial_t \bu + (\bu\bdot\grad)\bu = -\grad p + \nu\bigtriangleup \bu,
       \,\,\,\,\,\,\,\,\,\, \grad\bdot\bu=0 \lb{INS} \ee
where $\bx\in\Lambda\subset {\bf R}^d,$ for any integer $d\geq 2.$ Here
$p(\bx,t)$ is the so-called ``kinematic pressure'' (or, thermodynamically,
the enthalpy per unit mass). For any closed, oriented, rectifiable loop
$C\subset \Lambda $ at an initial time $t_0,$ one defines the circulation
\be \Gamma(C,t)=\oint_{C(t)}\bu(t)\bdot d\bx = \int_{S(t)}\bomega(t)\bdot
                            d{\bf A}  \lb{circ} \ee
where $C(t)$ is the loop at time $t$ advected by the fluid velocity, $S(t)$ is
any
surface spanning that loop,  and $\bomega(t)=\grad\btimes\bu(t)$ is the fluid
vorticity. These circulations satisfy the Kelvin-Helmholtz theorem in the
following sense:
\be {{d}\over{dt}}\Gamma(C,t)=\nu\oint_{C(t)}\bigtriangleup\bu(t)\bdot d\bx.
\lb{kelvin} \ee
%t.$
E.g., see \cite{Saffman1992}, section \S 1.6, for the standard derivation. It
is worth
observing that the Kelvin theorem for all loops $C$ is formally equivalent to
the Navier-Stokes equation \cite{Foiasetal2001}. Indeed, if  $\bu(\bx,t)$
is a smooth spacetime velocity field, divergence-free at all times $t$, then
equation (\ref{kelvin})  implies that
\be \oint_{C}\left[D_t\bu(t)-\nu\bigtriangleup\bu(t)\right]\bdot d\bx=0
\lb{K-INS}
\ee
for all loops $C$ at every time $t.$ Here
$D_t\bu=\partial_t\bu+(\bu\bdot\grad)\bu$
is the Lagrangian time-derivative and the equation (\ref{K-INS}) is derived by
applying (\ref{kelvin}) to the pre-image of the loop $C$ at initial time $t_0.$
By Stokes theorem, equation (\ref{K-INS}) can hold for all loops $C\subset
\Lambda$
if and only if there exists a pressure-field $p(\bx,t)$ such that the
Navier-Stokes
equation (\ref{INS}) holds locally and also globally, if the domain $\Lambda$
is simply connected.

In the inviscid limit $\nu\rightarrow 0,$ the circulation is formally conserved
for any initial loop $C.$ The fluid equations in this limit, the incompressible
Euler equations, are the equations of motion of a classical Hamiltonian
system. They can be derived by the Hamilton-Maupertuis principle from the
action functional
\be S[\bx] = {{1}\over{2}} \int_{t_0}^{t_f} dt \int_\Lambda  d\ba \,\,
|\dot{\bx}(\ba,t)|^2 \lb{action} \ee
with the pressure field $p(\bx,t)$ a Lagrange multiplier to enforce the
incompressibility constraint. Here $\bx(\ba,t)$ is the Lagrangian flow map
which satisfies $\dot{\bx}(\ba,t)=\bu(\bx(\ba,t),t)$ with initial condition
$\bx(\ba,t_0)
=\ba.$ See \cite{Salmon1988,Holmetal2002} for reviews. This variational
principle
yields the fluid equations in a Lagrangian formulation, as
$\ddot{\bx}(\ba,t)=-\grad
p(\bx(\ba,t),t).$ The Eulerian formulation (\ref{INS}) (with $\nu=0$) is
obtained
by performing variations in the inverse map $\ba(\bx,t),$ or ``back-to-labels
map'', with fixed particle positions $\bx.$ This Hamiltonian system has
an infinite-dimensional gauge symmetry group consisting of all
volume-preserving
diffeomorphisms of $\Lambda,$ which corresponds to all smooth choices
of initial fluid particle labels. In this framework, the conservation of the
circulations
for all closed loops $C$ emerges as a consequence of  Noether's theorem for
the particle-relabelling symmetry \cite{Arnold66}.  For reviews, see
\cite{Salmon1988},
Section 4 or \cite{Holmetal2002}, Section 2.2.

\subsection{Circulation and Turbulent Energy Dissipation}
\lb{taylor}

In several papers \cite{Taylor17,TaylorGreen37,Taylor1938}, G. I. Taylor
has argued for the importance of conservation of circulations in the turbulent
generation of energy dissipation at high Reynolds numbers in space dimension
$d=3$. We briefly review his ideas.  The simplest version of  Taylor's argument
is
based upon the concept of vortex line-stretching. Consider a vortex tube
initially with
length $L_0$, cross-sectional area $A_0,$ and vortex strength $\omega_0.$
Taylor
assumed that such a vortex tube at high Reynolds number  will evolve as a
material line.
Taylor also reasoned that vortex lines (or any material lines) should tend to
lengthen,
on average, under random advection by a turbulent velocity field. Thus, at a
later time
$t>t_0,$ the tube length is typically $L(t)>L_0.$ By incompressibility, the
volume $V(t)=
L(t)A(t)$ does not change in time, so that $A(t)<A_0.$ Furthermore, Taylor
reasoned  by
the Helmholtz theorem that the vortex-flux through the tube,
$\Gamma(t)=\omega(t)A(t),$
would not change, so that $\omega(t)>\omega_0.$ In fact, by this chain of
reasoning,
\be \omega(t)/\omega_0=L(t)/L_0 \lb{taylor} \ee
and vortex strength increases in direct proportion to line-length.  Because the
viscous
energy dissipation in the vortex-tube is given by $\nu \int \omega^2(t)
\,dV=\nu \omega^2(t)
V_0,$ this process should lead to a dramatic enhancement of dissipation.

However, this argument contains an apparent inconsistency. On the one hand,
Taylor's assumptions that vortex-lines are material lines and that the Kelvin
Theorem
applies require that the viscosity term in the circulation balance
(\ref{kelvin}) can be
neglected. On the other hand,  Taylor retains the viscous dissipation in the
energy
balance, arguing, in fact, that it is sizable. It is not at all clear that it
is valid to
ignore the viscosity effects in one place and to keep them in another. Taylor
himself recognized the delicacy of his argument. In  \cite{TaylorGreen37} he
presented this line of reasoning, and then wrote: ``When $\overline{\omega^2}$
has increased to some value which depends on the viscosity, it is no longer
possible
to neglect the effect of viscosity in the equation for the conservation of
circulation,
so that (10) [our (\ref{taylor})] ceases to be true.''  Thus, Taylor assumed
that there is
some interval of time or some range of length-scales for which viscous effects
can be
neglected in the circulation balance (\ref{kelvin}). We shall critically review
this
assumption below.

In a following paper \cite{Taylor1938}, Taylor tested some predictions of his
argument
using experimental data for decaying turbulence generated from a wind-tunnel.
His
analysis was based upon the following equation for production of enstrophy,
\begin{eqnarray}
 & & \partial_t\left({{1}\over{2}}|\bomega|^2\right)+
      \grad\bdot\left[{{1}\over{2}}|\bomega|^2\bu
      -\nu \grad\left({{1}\over{2}}|\bomega|^2\right)\right] \cr
   & & \,\,\,\,\,\,\,\,\,\,\,\,\,\,\,\,\,\,\,\,\,\,\,\,\,\,\,\,\,\,\,\,
       \,\,\,\,\,\,\,\,\,\,\,\,\,\,\,\,
   = \bomega^\top\bS\bomega
       -\nu|\grad\bomega|^2 \lb{3d-enstrophy}
\end{eqnarray}
which is an exact consequence of the incompressible Navier-Stokes dynamics
(\ref{INS}) for space-dimension $d=3.$ Here $S_{ij}=(1/2)(\partial u_i/\partial
x_j
+\partial u_j/\partial x_i)$ is the strain matrix. Under conditions of
space-homogeneity,
the average of the transport term vanishes, so that
\be (d/dt)\langle {{1}\over{2}}|\bomega|^2\rangle =
      \langle \bomega^\top\bS\bomega\rangle - \nu \langle
|\grad\bomega|^2\rangle
 \lb{mean-enstrophy} \ee
Taylor's argument on vortex-line stretching suggests that
$(d/dt)\langle|\bomega|^2\rangle
>0,$ which can hold if and only if
$ \langle \bomega^\top\bS\bomega\rangle >\nu \langle |\grad\bomega|^2\rangle
\geq 0 $
Thus, enstrophy will be created when the mean rate of vortex-stretching by the
strain
is positive and exceeds the mean destruction of enstrophy by viscosity. In
\cite{Taylor1938},
Taylor found from an analysis of wind-tunnel data that the latter condition
holds for an initial
range of time in decaying turbulence.

% \newpage

\section{Circulation Cascade}
\lb{circ_cascade}

We now turn to an analysis of circulation conservation in high-Reynolds-number
turbulent flow. One approach would be to directly analyze the $\nu\rightarrow
0$
limit of eq. (\ref{kelvin}). However, we shall pursue a complementary approach
based upon a study of nonlinear transfer in the inertial range.

\subsection{Filtering Approach}
\lb{filter}

To analyze the dynamics in the inertial range, we introduce effective equations
that govern the evolution of the velocity field at large length-scales. For any
chosen length $\ell,$ let
\be \Ol{\bu}_\ell(\bx)=\int d\br \,G_\ell(\br) \bu(\bx+\br) \lb{F-vel} \ee
denote the low-pass filtered velocity at scale $\ell,$ where $G_\ell(\br)=
\ell^{-d}G(\br/\ell)$ is a filter kernel. We shall assume that $G$ is positive,
smooth, rapidly decaying in space and with unit integral. Then $\Ol{\bu}_\ell$
satisfies an effective equation:
\be \partial_t\Ol{\bu}_\ell+(\Ol{\bu}_\ell\bdot \grad)\Ol{\bu}_\ell
+\grad\bdot\btau_\ell
   = -\grad\Ol{p}_\ell +\nu\bigtriangleup \Ol{\bu}_\ell
   , \lb{F-INS} \ee
where $\Ol{p}_\ell $ is the filtered pressure and $\btau_\ell$ is the turbulent
stress-tensor
\be \btau_\ell=\Ol{(\bu\,\bu)}_\ell-\Ol{\bu}_\ell\Ol{\bu}_\ell. \lb{stress} \ee
The filtering operation that we have employed can be regarded as a
``coarse-graining"
that eliminates high-wavenumber modes, as in renormalization-group methodology
\cite{Forsteretal77,Eyink94}. Because of momentum conservation, the effective
renormalized equation can change only by additional  contributions to the
stress
tensor. This filtering approach is also the mathematical basis of the
large-eddy-simulation
(LES) modeling scheme \cite{Germano92,MeneveauKatz00}. In this scheme, the
stress tensor is the main unknown which must be modelled, in order to obtain a
closed equation for computation of the large-scale velocity field.

% \newpage

In the inertial-range of turbulent flow the final viscosity term in
eq.(\ref{F-INS})
can be neglected. For example, a fairly crude estimate based upon the identity
\be \bigtriangleup \Ol{\bu}_\ell(\bx)=\ell^{-2}
                                  \int d\br \,(\bigtriangleup G)_\ell(\br)
\bu(\bx+\br) \lb{crude} \ee
is $\|\nu\bigtriangleup \Ol{\bu}_\ell\|_2\leq (\nu/\ell^2)({\rm
const.})\|\bu\|_2,$ where
$\|\bu\|_2=[\int d\bx\,|\bu(\bx)|^2]^{1/2}$ is the $L^2$-norm. If the total
kinetic energy per
mass $E=(1/2)\|\bu\|_2^2$ remains finite in the limit as $\nu\rightarrow 0,$
then the
viscosity term in eq.(\ref{F-INS}) tends to zero in $L^2$-norm for any fixed
filter-length $\ell.$

There is another form of the effective equation (\ref{F-INS}) which is useful.
Note that the
stress appears only via the {\it  turbulent (subgrid) force} $ \bF_\ell =
-\grad\bdot\btau_\ell$
\cite{MeneveauKatz00}. This can be replaced in (\ref{F-INS}) using the
following
elementary identity
\be  \bF_\ell = -\grad k_\ell + \bF_\ell^*, \lb{f-id} \ee
where $k_\ell=(1/2){\rm tr}\,\btau_\ell$ is the {\it turbulent kinetic energy}
\cite{Germano92} and
\be \bF_\ell^*= \Ol{(\bu\btimes\bomega)_\ell}
-\Ol{\bu}_\ell\btimes\Ol{\bomega}_\ell \lb{vort-force} \ee
is the {\it turbulent vortex force}. With this replacement, (\ref{F-INS})
becomes
\be \partial_t\Ol{\bu}_\ell+(\Ol{\bu}_\ell\bdot \grad)\Ol{\bu}_\ell =
-\grad\Ol{p}_\ell^* + \bF_\ell^* + \nu\bigtriangleup \Ol{\bu}_\ell, \lb{F-INS*}
\ee
where $\Ol{p}_\ell^*=\Ol{p}_\ell+k_\ell$ is a modified pressure. Although this
form
of the large-scale effective equation leads to more intuitive results, it is
less easy
to make sense of mathematically. In fact, the vortex force $\bF_\ell^*$ could
be
badly ultraviolet divergent in the limit as $\nu\rightarrow 0.$ Notice that
for infinite-Reynolds-number turbulence the velocity $\bu$ is believed to
be a continuous but non-differentiable function, so that the vorticity
$\bomega$
exists only as a distribution. Therefore, the product $\bu\btimes\bomega$ is
{\it a priori} ill-defined. However, the vortex force remains well-defined
due to the identity (\ref{f-id}), since both $\bF_\ell$ and $\grad k_\ell$
make sense as long as $\|\bu\|_2<\infty. $

\subsection{Circulation-Balance in the Large-Scales}
\lb{balance}

It is natural to inquire about the circulation-balance for the large-scale
effective equation. Let us choose an oriented, rectifiable, closed loop $C$
in space. We define $\Ol{C}_\ell(t)$ as the loop $C$ advected by the filtered
velocity  $\Ol{\bu}_\ell$. This definition makes sense, since the filtered
velocity
$\Ol{\bu}_\ell$ is Lipschitz in space, and corresponding flow maps
$\Ol{\bx}_\ell(\ba,t)$ defined by
 \be (d/dt)\Ol{\bx}_\ell(\ba,t)=\Ol{\bu}_\ell(\Ol{\bx}_\ell(\ba,t),t),\,\,\,\,
        \Ol{\bx}_\ell(\ba,t_0)=\ba, \lb{lag-map} \ee
both exist and are unique (see I).  We define a ``large-scale circulation''
with initial loop $C$ as the line-integral
\be \Ol{\Gamma}_\ell(C,t) = \oint_{\Ol{C}_\ell(t)}
                               \Ol{\bu}_\ell(t)\cdot d\bx. \lb{F-circ} \ee
for $\ell<R=$ the radius of gyration of the loop $C$ \footnote{The circulation
quantity defined in (\ref{F-circ}) is mainly of interest for $\ell < R$.
Otherwise, if
$\ell\gg R,$ then $\Ol{\bu}_\ell$ is nearly constant over the scale of the loop
and $|\Ol{\Gamma}_\ell(C,t)|=O(R/\ell)\ll 1.$}. The same calculation that
establishes
the Kelvin theorem, but using the effective eq. (\ref{F-INS}) rather than
Navier-Stokes eq.(\ref{INS}), gives
\be
(d/dt)\Ol{\Gamma}_\ell(C,t) = \oint_{\Ol{C}_\ell(t)}[\bF_\ell(t)+
          \nu\bigtriangleup\Ol{\bu}_\ell] \cdot d\bx
%                                                =
%%\oint_{\Ol{C}_\ell(t)}\bF_\ell^*(t)\cdot d\bx
% \Ol{\Gamma}_\ell(C,t)-\Ol{\Gamma}_\ell(C,t_0)= \int_{t_0}^t d\tau\,
% \oint_{\Ol{C}_\ell(\tau)}\bF_\ell(\tau)\cdot d\bx.
\lb{F-kelvin} \ee
If the Navier-Stokes eq.(\ref{INS}) were driven by an external body-force
$\bF^{{\rm ext}},$ then there would be an additional term $\Ol{\bF}^{{\rm
ext}}$
inside the square bracket in eq.(\ref{F-kelvin}). If this external force is
spectrally
supported at wavenumbers of order $1/L,$ then its contribution  to the
circulation balance is $O(R/L).$  Thus, the forcing term is negligible for
$R\ll L.$
Likewise, the viscous term in eq.(\ref{F-kelvin}) is negligible  for small
viscosity
$\nu$ and fixed filter-length $\ell,$ by an elaboration of the argument given
around eq.(\ref{crude}). (A so-called ``trace theorem''  can be used to
estimate the restriction of $\bigtriangleup\Ol{\bu}_\ell$ to the loop $C$; see
I and \cite{Triebel1983}).
% \footnote{
% $\|\bigtriangleup\Ol{\bu}_\ell \|_{H^{s-2}_2({\bf T}^d)} \leq ({\rm
%%const.})\ell^{-s}\|\bu\|_{L^2({\bf T}^d)}$.
% By trace theorems,
% $$\|\bigtriangleup\Ol{\bu}_\ell\|_{H^{s-2-(d-1)/2}_2(C)}\leq ({\rm const.})
% \|\bigtriangleup\Ol{\bu}_\ell\|_{H^{s-2}_2({\bf T}^d)}$$
% Thus, for $s>(d+3)/2,$
% $$ \|\bigtriangleup\Ol{\bu}_\ell\|_{L^2(C)} \leq ({\rm
%%const.})\ell^{-s}\|\bu\|_{L^2({\bf T}^d)} $$}

These remarks show that the nonlinear term from the subgrid force is the
dominant
term in the circulation balance (\ref{F-kelvin}) for inertial-range values
$L\gg R>\ell
\gg \eta_d$ (where $\eta_d$ is a dissipation length-scale determined by the
viscosity
$\nu$). If we imagine that the total circulation at all scales on the loop is
conserved,
then the line-integral of $\bF_\ell$ on the RHS of (\ref{F-kelvin}) represents
a ``transfer''
of circulation to subgrid modes at length-scales $<\ell$. This motivates the
definition,
for any loop $C$ and filter length $\ell,$ of a {\it flux of circulation}
\be K_\ell(C,t)= -\oint_{\Ol{C}_\ell(t)} \bF_\ell(t)  \cdot d\bx
                        = -\oint_{\Ol{C}_\ell(t)} \bF_\ell^*(t)  \cdot d\bx
\lb{circ-flux} \ee
so that $(d/dt)\Ol{\Gamma}_\ell(C,t)=-K_\ell(C,t)$ (up to small corrections
from
external forcing and viscosity). We have used identity (\ref{f-id}) to justify
the equality of the two expressions in the definition (\ref{circ-flux}). The
minus
sign has been introduced so that the signs of the circulation (\ref{F-circ})
and
the circulation-flux (\ref{circ-flux}) should be positively correlated. This
expectation will be discussed more below.

The ``circulation-flux'' defined in (\ref{circ-flux}) has the physical
dimensions
of work or of torque (per unit mass). Additional insight into its meaning can
be
obtained by decomposing the turbulent vortex-force (\ref{vort-force}) into
components perpendicular and parallel to large-scale vortex-lines:
\be  \bF_{\perp\,\ell}^*  = \bsigma_\ell\btimes\hat{\bomega}_\ell,
\,\,\,\,\,\,\,\,
       \bF_{\|\,\ell}^* = (\bF_\ell^*\bdot\hat{\bomega}_\ell)\hat{\bomega}_\ell
       \lb{vort-force-dec} \ee
where $\hat{\bomega}_\ell=\Ol{\bomega}_\ell/|\Ol{\bomega}_\ell|$ and
\be  \bsigma_\ell =\hat{\bomega}_\ell\btimes\bF_\ell^*. \lb{sigma} \ee
If $\hT_\ell$ is the unit tangent vector to the curve $\Ol{C}_\ell(t)$ and $s$
is the arc-length
parameter, then
\be K_\ell(C,t)= \oint_{\Ol{C}_\ell(t)} \bsigma_\ell(t) \bdot \bn_\ell \,ds
                        - \oint_{\Ol{C}_\ell(t)} \bF_{\|\,\ell}^*(t)  \cdot
d\bx
               \lb{circ-flux-dec} \ee
where $\bn_\ell=\hT_\ell\btimes\hat{\bomega}_\ell.$ Note that the latter vector
is normal
both to lines of large-scale vorticity $\Ol{\bomega}_\ell $ and to the loop
$\Ol{C}_\ell(t),$
but it is not generally a unit vector. The first term in (\ref{circ-flux-dec})
can be interpreted
as a lateral diffusion of vortex-lines out of the advected loop,  where
$\bsigma_\ell$
plays the role of a transport vector of vortex-lines. The second term in
(\ref{circ-flux-dec})
represents an additional work (or torque) due to the parallel component of the
turbulent
vortex-force.

Some particular cases of (\ref{circ-flux-dec}) are of special interest. For
example,
consider the case  that $\Ol{C}_\ell(t)$ is instantaneously a closed vortex
line. (This
property will not generally be preserved in time). Then the first term in
(\ref{circ-flux-dec})
vanishes and $K_\ell(C,t)= - \oint_{\Ol{C}_\ell(t)} \bF_{\|\,\ell}^*(t)  \cdot
d\bx.$ Such integrals
play an important role in vortex-reconnection theory \cite{Hornig01}. The
distinguished
vortex lines for which this integral is extremal drive the reconnection process
and the
value of the integral for such lines gives the rate of reconnection of
vortex-flux. This
integral is therefore the proper point of departure for a theory of turbulent
reconnection
of large-scale vortex-lines. Another special case of  (\ref{circ-flux-dec}) of
interest is
when the loop $\Ol{C}_\ell(t)$ lies in a transversal surface normal to the
lines of
large-scale vorticity. In that case, the second term in (\ref{circ-flux-dec})
vanishes
and $ K_\ell(C,t)= \oint_{\Ol{C}_\ell(t)} \bsigma_\ell(t) \bdot \hn_\ell \,ds,
$ where
$\hn_\ell=\hT_\ell\btimes\hat{\bomega}_\ell$ is now a unit vector. This
condition
is always satisfied for space dimension $d=2.$ The flux of circulation is then
entirely
due to the diffusion of vortex-lines out of the loop.

These remarks on physical interpretation of $K_\ell(C,t)$ lead to some natural
guesses on the correlation of its sign with that of the circulation
$\Ol{\Gamma}_\ell(C,t).$
The latter can be written as
\be \Ol{\Gamma}_\ell(C,t) = \int_{\Ol{S}_\ell(t)}
                               \Ol{\bomega}_\ell(t)\bdot d\bA, \lb{F-helmholtz}
\ee
where $\Ol{S}_\ell(t)$ is any smooth surface spanning the loop $\Ol{C}_\ell(t)$
and with
orientation consistent to that of $\Ol{C}_\ell(t)$ (by the righthand rule).  If
the circulation
(\ref{F-helmholtz}) is positive, then there is a net contribution from vortex
lines threading
the loop in the direction of the surface unit normal. If the effect of the
turbulence is ``diffusive"
on average, then one would expect that the vortex-force will tend to smooth out
the excess
of positive-sign vorticity threading the loop. Thus, according to the sign
convention of
the definition (\ref{circ-flux}), we can expect that $K_\ell(C,t)$ will also
tend to be positive
and to reduce the overall magnitude of the large-scale circulation. Of course,
this argument
works equally well when $\Ol{\Gamma}_\ell(C,t)$ has negative sign. We may
therefore
expect that there is in general a ``forward cascade'' of circulations, and that
the magnitude
of the large-scale circulation, of whatever sign, will tend to be decreased by
the small-scale
turbulence. This reasonable result has been confirmed by numerical results in
\cite{Chenetal06}.

An interesting exception is the inverse-energy cascade for $d=2 $ turbulence.
For space dimension $d=2,$ the enstrophy $\Omega(t)=(1/2)|\omega|^2$ is
an inviscid invariant and its flux to unresolved scales $<\ell$ is measured by
\be   Z_\ell = -\grad\Ol{\omega}_\ell\bdot \bsigma_\ell, \lb{Z-flux} \ee
where $\Ol{\omega}_\ell$ is the filtered vorticity (perpendicular to the plane)
and
$\bsigma_\ell$ is the vorticity transport vector defined in (\ref{sigma}).
See \cite{Eyink96,Eyink00,Chenetal03}. From (\ref{Z-flux}) one can see that
enstrophy will cascade forward to small scales when vorticity transport
tends to be ``down-gradient" and $\grad\Ol{\omega}_\ell\bdot \bsigma_\ell<0.$
On the other hand, enstrophy flux will be inverse to large-scales when the
vorticity transport is ``up-gradient.''  In $d=2$ there are expected to be two
inertial
cascade ranges, the direct enstrophy cascade where the mean enstrophy
flux is positive and the inverse energy cascade where the mean energy flux
is negative \cite{Kraichnan67,Batchelor69}. However, there is also some
``leakage'' of energy flux and enstrophy flux into the opposite ranges
(e.g. see \cite{Borue93,Eyink96}). In particular, the mean enstrophy flux
in the inverse energy cascade range is {\it negative}, or toward larger scales.
This means, according to (\ref{Z-flux}), that the vorticity transport in that
range
is, on average, ``up-gradient'' or ``anti-diffusive''. Therefore, our argument
for the sign of circulation-flux is reversed. In the inverse cascade range,
a loop containing an excess of one sign of vorticity should tend to accumulate
more vorticity of the same sign. Thus, in the $d=2$ inverse energy cascade
range there should be also an ``inverse cascade of circulations''
\footnote{This phenomenon has been previously observed in numerical
simulations of the $d=2$ inverse energy cascade (Minping Wan and Shiyi Chen,
private communication).}.

\subsection{\bf Stretching of Large-Scale Vorticity}
\lb{large_stretch}

We have seen that the ``large-scale circulations'', in the inertial range,
evolve
according to the equation
\be (d/dt)\Ol{\Gamma}_\ell(C,t) = \oint_{\Ol{C}_\ell(t)} \bF_\ell^*(t)\cdot
d\bx.
      \lb{F-kelvin*} \ee
The term on the righthand side due to the vortex-force need not be negligible.
Thus, Taylor's conjecture that Kelvin's theorem should hold in the inertial
range,
even approximately, is far from obviously true. In the next section we shall
explore this question mathematically, to the extent possible. Here we
discuss some physical implications of Taylor's conjecture, if true.

If we suppose that the inertial-range circulations are conserved, then Taylor's
argument about vortex line-stretching can be repeated for filtered vorticity,
implying
\be  (d/dt)\langle |\Ol{\bomega}_\ell|^2\rangle >0. \lb{L-stretch} \ee
This result can also be understood from the equation for the filtered
vorticity,
obtained by taking the curl of  equation (\ref{F-INS*}) (with $\nu=0$):
\be \partial_t\Ol{\bomega}_\ell+(\Ol{\bu}_\ell\bdot \grad)\Ol{\bomega}_\ell
    = (\Ol{\bomega}_\ell\bdot \grad)\Ol{\bu}_\ell +\grad\btimes\bF_\ell^*.
    \lb{F-omega-eq} \ee
{}From this an equation for inertial-range enstrophy easily follows:
\begin{eqnarray}
& & \partial_t\left({{1}\over{2}}|\Ol{\bomega}_\ell|^2\right)+
      \grad\bdot\left[{{1}\over{2}}|\Ol{\bomega}_\ell|^2\Ol{\bu}_\ell
      +|\Ol{\bomega}_\ell|\bsigma_\ell\right] \cr
& & \,\,\,\,\,\,\,\,\,\,\,\,\,\,\,\,\,\,\,\,\,\,\,\,\,\,\,\,\,\,\,\,
    = \Ol{\bomega}_\ell^\top\Ol{\bS}_\ell\Ol{\bomega}_\ell
    +\bF_\ell^*\bdot(\grad\btimes\Ol{\bomega}_\ell).   \lb{F-enstrophy}
\end{eqnarray}
[Compare with \cite{Meneveau94}, eq.(51), for $\nu\rightarrow 0.$] Notice
that the vorticity transport vector $\bsigma_\ell$ defined in (\ref{sigma})
contributes to the space transport of enstrophy. However, assuming
space-homogeneity, all of the space-flux terms average to zero and
\be (d/dt)\langle{{1}\over{2}}|\Ol{\bomega}_\ell|^2\rangle
    = \langle\Ol{\bomega}_\ell^\top\Ol{\bS}_\ell\Ol{\bomega}_\ell\rangle
    +\langle\bF_\ell^*\bdot(\grad\btimes\Ol{\bomega}_\ell)\rangle.
    \lb{avrg-F-enstrophy} \ee
This equation is an exact inertial-range analogue of equation
(\ref{mean-enstrophy}) for total enstrophy. The first term on the
righthand side of (\ref{avrg-F-enstrophy}) represents inertial-range
vortex-stretching and the second term represents enstrophy flux
to length-scales $<\ell.$ For freely decaying turbulence at
early times, Taylor's argument predicts that $\langle
\Ol{\bomega}_\ell^\top\Ol{\bS}_\ell\Ol{\bomega}_\ell\rangle +
\langle\bF_\ell^*\bdot(\grad\btimes\Ol{\bomega}_\ell)\rangle
=(d/dt)\langle{{1}\over{2}}|\Ol{\bomega}_\ell|^2\rangle>0.$
On physical grounds, one expects that the vortex-stretching is
positive and the enstrophy transfer term negative, with the net enstrophy
production positive. At later times a quasi-equilibrium should be
established so that $(d/dt)\langle{{1}\over{2}}|\Ol{\bomega}_\ell|^2
\rangle\approx 0$ and the dominant balance becomes
\be 0< \langle\Ol{\bomega}_\ell^\top\Ol{\bS}_\ell\Ol{\bomega}_\ell\rangle
   \approx -\langle\bF_\ell^*\bdot(\grad\btimes\Ol{\bomega}_\ell)\rangle
   \lb{pos-stretch} \ee
For some experimental results on these questions, see \cite{Meneveau94}.

It was observed in \cite{BorueOrszag98} that the energy flux $\Pi_\ell$
to unresolved scales $<\ell$ can be expressed approximately in terms
of the negative skewness of filtered strain and the stretching rate of
filtered vorticity:
\be
\Pi_\ell = C\ell^2\left[-{\rm tr}\,\left(\Ol{{\bf S}}_\ell^3\right)
 +(1/4)\Ol{\bomega}_\ell^\top\Ol{\bS}_\ell\Ol{\bomega}_\ell\right].
 \lb{BO-energy-flux} \ee
This expression is the first term in a systematic ``multi-scale gradient
expansion" \cite{Eyink06a}. It follows from an identity of Betchov
\cite{Betchov56} that for any homogeneous turbulence
\be \langle\Pi_\ell\rangle = C\ell^2\langle\Ol{\bomega}_\ell^\top
      \Ol{\bS}_\ell\Ol{\bomega}_\ell\rangle. \lb{Betchov} \ee
Thus, the energy cascade will be forward to small scales when
the mean rate of vortex-stretching is positive. This is an inertial-range
version of Taylor's mechanism \cite{TaylorGreen37,Taylor1938}.

\subsection{Anomalous Conservation of Circulation}
\lb{anomaly}

We now consider the question whether Kelvin's Theorem can hold,
in any sense, in turbulent flow at high Reynolds number. In view
of equation (\ref{F-kelvin}) or (\ref{F-kelvin*}), we must estimate the
magnitude of the circulation-flux defined in (\ref{circ-flux}). The following
simple identity, observed in \cite{Eyink06}, is useful to provide an estimate
of the turbulent subgrid force:
\begin{gather}
f_{\ell\,i}(\bx) = \frac{1}{\ell}\int d\br \,(\partial_j G)_\ell(\br) \,\delta
u_i(\br;\bx)
    \delta u_j(\br;\bx) \lb{f-delta} \\
-\frac{1}{\ell}\int d\br \,(\partial_j G)_\ell(\br) \,\delta u_i(\br;\bx)
        \int d\br' \,G_\ell(\br') \,\delta u_j(\br';\bx). \notag
\end{gather}
%
%
% \begin{eqnarray}
% & & f_{\ell\,i}(\bx) = \frac{1}{\ell}\int d\br \,(\partial_j G)_\ell(\br)
%%\,\delta u_i(\br;\bx)
%    \delta u_j(\br;\bx) \lb{f-delta} \cr
% & &
%   -\frac{1}{\ell}\int d\br \,(\partial_j G)_\ell(\br) \,\delta u_i(\br;\bx)
%        \int d\br' \,G_\ell(\br') \,\delta u_j(\br';\bx).
% \end{eqnarray}
%
Here $\delta\bu(\br;\bx)=\bu(\bx+\br)-\bu(\bx)$ is the velocity-increment with
separation vector $\br$ at location $\bx.$ An upper bound easily follows
that $|\bF_\ell|=O(|\delta u(\ell)|^2/\ell),$ where $\delta u(\ell)$ is the
maximum
magnitude of the velocity-increment for separation vectors with $|\br|<\ell$
\cite{Eyink06}.

If the velocity field were smooth, then $|\delta u(\ell)|\sim ({\rm
const.})\ell$
for small $\ell$ and the subscale force would vanish as $\ell\rightarrow 0$.
However, a turbulent velocity field does not remain smooth in the limit
as the Reynolds number tends to infinity. Instead, theory, simulations,
and experiment indicate that the velocity field is only H\"{o}lder continuous
with exponent $0<h<1:$
\be   |\delta\bu(\br;\bx)|=O(r^h). \lb{hoelder} \ee
At each point $\bx$ one refers to the maximal value $h$ for which
(\ref{hoelder})
holds as the {\it H\"{o}lder exponent} at that point. There is a spectrum of
such
singularities in the flow, with exponent $h$ occurring on a set ${\mathcal
S}(h)$
with fractal dimension $D(h).$  It was pointed out by Onsager
\cite{Onsager1949}
that the smallest exponent $h_{\min}$ must be $\leq 1/3$ to explain
non-vanishing
energy dissipation in the inviscid limit. Parisi and Frisch
\cite{FrischParisi85} invoked
a multifractal spectrum $D(h)$ of singularities to explain the anomalous
scaling of
$p$th moments of velocity-increments (so-called $p$th-order
structure-functions).
Such multifractal spectra of H\"{o}lder exponents have been confirmed by
analysis
of data from experiments and simulations  \cite{Muzyetal91,Arneodoetal95,
KestenerArneodo04}. Of course, at finite Reynolds numbers there are only
``near-singularities'' in the inertial-range of scales and the velocity is
smooth
in the dissipation range, where effects of viscosity are important.

{}From our estimate below eq.(\ref{f-delta}), we see that $|\bF_\ell |
=O(\ell^{2h-1})$
at any point with local H\"{o}lder exponent $h$.  Thus, the circulation flux
$K_\ell(C,t)$
will go to zero as $\ell\rightarrow 0$ if the smallest velocity H\"{o}lder
exponent
$h_{\rm min}$ is $>1/2$ and if also the curve $C(t)$ has finite length
\cite{Eyink06}.
This is an exact analogue for circulation flux of Onsager's result
\cite{Onsager1949}
for vanishing of energy flux when $h_{\rm min}>1/3.$  Only a sufficiently rough
velocity field can provide a transport of vortex lines which is non-vanishing
in the
limit as $\ell\rightarrow 0.$  However, high Reynolds turbulence in space
dimension
$d=3$ has a plethora of singularities with exponents $h\leq 1/2.$  For example,
the most probable exponent $h_*$ with $D(h_*)=3$ has a value $h_*\doteq 1/3,$
very close to the mean-field Kolmogorov value \cite{Muzyetal91,Arneodoetal95,
KestenerArneodo04}. Furthermore, the curves $\Ol{C}_\ell(t)$ advected by the
large-scale velocity $\Ol{\bu}_\ell$ are expected to approach a fractal curve
$C(t)$ in the limit as $\ell\rightarrow 0$
\cite{Mandelbrot76,SreenivasanMeneveau1986}.
Thus, circulation-flux is not likely to vanish as the filtering length
decreases
through the inertial-range. Numerical simulations of high-Reynolds-number
turbulence for $d=3$ confirm this prediction \cite{Chenetal06}.

There is an important subtlety in the formulation of Kelvin's theorem for
infinite-Reynolds-number turbulence that must be mentioned at this
point. Recent work on an idealized turbulence problem---the Kraichnan
model of random advection \cite{Kraichnan68}---has shown that Lagrangian
particle trajectories $\bx(t),\,\bx'(t)$ can explosively separate even
when $\bx_0=\bx'_0$ initially, if the advecting velocity field is only
H\"{o}lder continuous and not Lipschitz. See \cite{Bernardetal1998}.
Mathematically, this is a consequence of the non-uniqueness of solutions
to the initial-value problem, while, physically, it corresponds to the
two-particle
turbulent diffusion of Richardson \cite{Richardson1926}. It has been rigorously
proved in \cite{LeJanRaimond2002,LeJanRaimond2004} that there is a random
process of Lagrangian particle paths $\bx(t)$ in the Kraichnan model for a
fixed
realization of the advecting velocity and a fixed initial particle position.
This phenomenon
has been termed {\it spontaneous stochasticity} \cite{Chavesetal2003} and it
is likely that it holds,  not only in the Kraichnan model, but also for
singular solutions
of the inviscid Euler equations. If so, then the advected curves $C(t)$ that
appear
in the definition of circulation (\ref{circ}) are likely to be {\it random}
fractal curves!

If these speculations are correct, then the time-series of circulations
$\Gamma(C,t)$
are also a stochastic process, for a fixed turbulent velocity field.  In
\cite{Eyink06}
we have presented some plausibility arguments in favor of the following
``martingale property" for this random process of circulations:
\be
\langle\Gamma(C,t)|\Gamma(C,\tau),\tau<t'\rangle=\Gamma(C,t'),\,\,\,\mbox{for
$t>t'.$}
\lb{martingale} \ee
Here $\langle\cdot\rangle$ denotes the expectation over the ensemble of random
Lagrangian paths and we have conditioned on the past circulation history
$\{\Gamma(C,\tau),\tau<t'\}. $ Heuristically,
\begin{eqnarray}
& & (d/dt)\langle\Gamma(C,t)|\Gamma(C,\tau),\tau<t'\rangle= \cr
& & \,\,\,\,\,\,\,\,\,\,\,\,\,\,\,\,\,\,\,\,
-\lim_{\ell\rightarrow 0}\langle K_\ell(C,t)|\Gamma(C,\tau),\tau<t'\rangle
 =0. \lb{dot-circ}
\end{eqnarray}
The circulation-flux in (\ref{dot-circ}) is conjectured to average to zero,
due to increasingly rapid oscillations of the vortex-force $\bF_\ell^*$
around the loop $\Ol{C}_\ell(t),$ as $\ell\rightarrow 0.$ See
\cite{Eyink06}. The result in (\ref{dot-circ}) has been partially confirmed
by the results of a numerical simulation in \cite{Chenetal06}, providing
some support to the conjecture (\ref{martingale}).  This ``martingale
property"
is a statement of conservation of circulations, in a conditional mean sense.
It is not clear yet whether this weakened version of the Kelvin theorem
is valid and, if so, whether it suffices for Taylor's vortex-stretching
mechanism.

\subsection{Analogy with Phase-Slip in Superfluids}
\lb{phase-slip}

It is worth pointing out an analogy of the ``circulation cascade'' discussed
above with another physical phenomenon, the ``phase-slip'' due to quantized
vortex lines in superfluids \cite{Anderson66,Donnelly91}. Anderson had
already discussed classical analogues of quantum phase-slip in
\cite{Anderson66},
Appendix B. His starting point was the classical Euler equations for an
incompressible
fluid, written as
\be \partial_t\bu=-\grad h+\bu\btimes\bomega, \lb{Euler} \ee
where $h=p+(1/2)|\bu|^2$ is the enthalpy. Anderson considered the line-integral
of the fluid velocity $\bu$ along a {\it stationary} curve $C$ connecting two
points
$P_1$ and $P_2,$ showing that
\be (d/dt)\int_C\,\bu(t)\bdot d\bx = -\Delta_{{\,\!}_C} h
           + \int_C \,(d\bx\btimes \bu)\bdot\bomega.  \lb{phase-slip} \ee
Here $\Delta_{{\,\!}_C} h=h(P_2)-h(P_1)$ is the difference of $h$ along the
curve $C.$ Denoting time-average by $\overline{(\cdot)},$ this relation yields
\be \Delta_{{\,\!}_C} \overline{h}= \int_C \,\overline{(d\bx\btimes
\bu)\bdot\bomega}.
\lb{phase-slip2} \ee
Since vortex lines for smooth solutions of the classical Euler equations move
with the particle velocity $\bu=d\bx/dt,$ the righthand side of
(\ref{phase-slip2})
can be interpreted as an average rate of flow of vorticity across the curve
$C.$ This
flow rate is thus equal to the  average enthalpy difference along the curve.
After deriving (\ref{phase-slip2}), Anderson wrote \cite{Anderson66}: ``We see
immediately that this equation is far more important in a superfluid, where
vorticity
is conserved and quantized, than it is in ordinary fluids, where in a laminar
flow,
for instance, the right-hand side has little or no special significance.'' One
critical
difference between classical fluids and superfluids is that, in the former, the
vortex-lines for laminar solutions move with the fluid.  Thus, if one instead
considers
a {\it material} curve $C(t)$,  advected by the fluid velocity $\bu,$ then one
obtains
\be (d/dt)\int_{C(t)}\,\bu(t)\bdot d\bx = \Delta_{{\,\!}_{C(t)}} \lambda
\lb{half-kelvin} \ee
with $\lambda=(1/2)|\bu|^2-p,$ rather than (\ref{phase-slip}). The nontrivial
term
associated to flow of vorticity across the curve is now absent and
eq.(\ref{half-kelvin})
for a closed loop yields the classical Kelvin Theorem.

Nevertheless, we have found that it is possible for turbulent flow to yield a
nontrivial
result. In fact, by filtering the Euler equation (\ref{Euler}) one obtains
\be \partial_t\Ol{\bu}_\ell
=-\grad \Ol{h}_\ell+\Ol{\bu}_\ell\btimes\Ol{\bomega}_\ell+\bF^*_\ell,
\lb{F-Euler} \ee
with the additional vortex-force term. This equation is equivalent to
\be \Ol{D}_t\Ol{\bu}_\ell=\partial_t\Ol{\bu}_\ell+(\Ol{\bu}_\ell\bdot
\grad)\Ol{\bu}_\ell
                                = -\grad\Ol{p}_\ell^* + \bF_\ell^*,
\lb{F-Euler*} \ee
which is our old eq.(\ref{F-INS*}) for $\nu=0.$ As we have seen in our earlier
discussion of the large-scale circulation balance, eq.(\ref{F-kelvin}) or
(\ref{F-kelvin*}),
the turbulent vortex-force provides a nontrivial transport of vorticity across
material
curves. Here it is crucial that the velocity field be sufficiently singular, to
permit
a transport which is non-vanishing for $\ell\rightarrow 0.$ If instead the flow
were
smooth and laminar, then $\bF_\ell^*\rightarrow 0$ in that limit and filtering
the
equation would lead to no new result. For singular solutions the Euler equation
(\ref{Euler}) {\it must} be filtered to make sense, as a matter of principle.
In the
presence of singularities the equation is interpreted in the sense of
distributions,
which means that it must be smeared with smooth test functions.

Nontrivial results are also possible in superfluids, for similar reasons. The
superfluid
phase order parameter $\varphi$ obeys  the Josephson-Anderson frequency
equation
\cite{Anderson66,Donnelly91}:
\be \hbar \,d\varphi/dt=-(\mu+\frac{1}{2}mu^2_s), \lb{phi-dyn} \ee
where $\mu$ is the chemical potential and $\bu_s=(\hbar/m)\grad\varphi$
is the superfluid velocity. It is straightforward to derive from
(\ref{phi-dyn}) the
superfluid equation of motion
\be D_t \bu_s = (\partial_t+\bu_s\bdot\grad)\bu_s=
                         -\grad(\mu/m)-\bu_s\btimes\bomega_s.  \lb{super-euler}
\ee
Here the final term contains the superfluid vorticity
$\bomega_s=\grad\btimes\bu_s$
which is, formally, a delta-function supported on singular vortex lines (zeroes
of the superfluid  density).  Equation (\ref{super-euler}) is the basis of
derivations
of Kelvin's theorem for superfluids, e.g. see \cite{DamskiSacha03} in the
context
of the zero-temperature Gross-Pitaevskii equation.  Note, however, that such
derivations require that the advected loop not pass through singular points
where
the superfluid velocity is ill-defined. Since the quantized vortex lines are
{\it not}
material lines in general (e.g. see
\cite{Thoulessetal99,AoZhu99,Nilsenetal05}),
it is possible for them to migrate out of an advected loop. Examples are given
in \cite{DamskiSacha03} of the failure of Kelvin's theorem due to the
intersection
of loops with singularities that are, formally, represented by the rightmost
term
in eq.(\ref{super-euler}). That equation is thus analogous to
eq.(\ref{F-Euler*})
for classical turbulence.

One of the concrete manifestations of quantum phase slip is the decay of
``persistent'' superfluid flow in a thin toroidal ring. E.g. see
\cite{Muelleretal98}
and references therein. This process has a number of similarities to the
``cascade
of circulations'' in turbulent flow. The decay of the superflow is mediated by
the
(thermal or quantum) nucleation of quantized vortices which migrate out of
the ring. The passage of a vortex across the toroidal cross-section induces
by phase-slip a pulse of torque which decreases the circulation around the
ring.
The reduction in the angular momentum of the superfluid condensate is balanced
by a gain in the normal fluid excitations, acting as an angular momentum
reservoir.
In the turbulent circulation-cascade, the large-scale vortex lines are also not
material,
because singularities in the velocity field allow them to diffuse relative to
the fluid.
The subscale modes at length-scales $<\ell$ act as a reservoir, whose feedback
on the resolves scales $>\ell$ provides the vortex-force that drives the
diffusion.
Unlike in superfluids, this is a continuous process, since classical vortices
are not
quantized. There is also no need for the singularities to be nucleated as
fluctuations,
since they are everywhere present in the turbulent flow. Finally, if the
``martingale''
conjecture (\ref{martingale}) is correct, then the turbulent diffusion of
vortex-lines
is not persistent in scale, on average, and does not lead to irreversible mean
decay
of circulations.

\subsection{Scale-Locality and MSG Expansion}
\lb{locality}

We have referred to this turbulent diffusion of vorticity as a ``cascade'' of
circulations,
but we have not shown that the process is a local-in-scale cascade. Here we
shall
examine this issue, following the general approach in \cite{Eyink05}.

We note first that the turbulent vortex-force $\bF_\ell^*$ defined in
(\ref{vort-force})
is {\it a priori} not ultraviolet (UV)-local, under conditions realistic for
turbulence
in $d=3$. In fact, the vorticity is a dissipation-range variable and its
largest
contributions come from the viscous scale. The arguments in \cite{Eyink05}
for UV-locality would apply to $\bF_\ell^*$ if the H\"{o}lder exponents $h_u$
of velocity and $h_\omega$ of vorticity {\it both} were positive. However,
$h_\omega
=h_u-1,$ so that vorticity is expected to have negative H\"{o}lder exponents
in the infinite-Reynolds-number limit (and thus to exist only as a
distribution)
\cite{KestenerArneodo04}. It is possible that there could be cancellations in
the
average (\ref{vort-force}) over displacement vectors that defines the
vortex-force.
E.g. this was found to be true in the $d=2$ enstrophy cascade, by an analysis
of the
results of a numerical simulation \cite{Chenetal03}. However, the UV-divergence
is more severe for $d=3$, so that sufficient cancellation is less likely there.

On the other hand, because of the identity (\ref{f-id}), we may use the
turbulent
subscale force $\bF_\ell=-\grad\bdot\btau_\ell$ rather than the vortex-force
$\bF_\ell^*$
to study the circulation-flux. The force $\bF_\ell$ has much greater chance to
be
scale-local, because it is defined only in terms of velocity. Indeed, some
locality
properties follow directly from the representation (\ref{f-delta}) in terms of
velocity
increments. As in \cite{Eyink05}, let us define $\bu^{>\Delta}=G_\Delta*\bu$ to
be
the low-pass filtered velocity at length-scale $\Delta>\ell$ and define
$\bu^{<\delta}
=\bu-\bu^{>\delta}$ to be the high-pass filtered velocity at length-scale
$\delta<\ell.$
We can then define a very large-scale contribution $\bF_\ell^{>\Delta}$ to the
turbulent force by replacing both $\bu$ in the formula (\ref{f-delta}) with
$\bu^{>\Delta}.$
Likewise, we define a very small-scale contribution $\bF_\ell^{<\delta}$ by
replacing
both $\bu$ with $\bu^{<\delta}.$ Now suppose that the velocity field has
H\"{o}lder
exponent $h$ at a considered point $\bx.$ Then, the following estimates can be
easily derived, by the same methods as in \cite{Eyink05}:
\be |\bF_\ell^{>\Delta}|=O\left(\ell \Delta^{2h-2}\right) \lb{IR-local} \ee
and
\be |\bF_\ell^{<\delta}|=O\left(\delta^{2h}/\ell\right). \lb{UV-local} \ee
The estimate (\ref{IR-local}) expresses infrared (IR)-locality. In fact,
when $h<1,$ this estimate shows that $\bF_\ell^{>\Delta}$ decreases
for increasing $\Delta$ and fixed $\ell$. Relative to the estimate
$|\bF_\ell|=O(\ell^{2h-1}),$ the estimate (\ref{IR-local}) for
$|\bF_\ell^{>\Delta}|$
is smaller by a factor $O((\ell/\Delta)^{2(1-h)}).$ Likewise, the estimate
(\ref{UV-local}) expresses UV-locality. When $h>0,$ this estimate shows
that $\bF_\ell^{<\delta}$ decreases for decreasing $\delta$ and fixed $\ell$.
The estimate (\ref{UV-local}) for $|\bF_\ell^{<\delta}|$ is smaller than that
for $|\bF_\ell|$ by a factor of $O((\delta/\ell)^{2h}).$ These results show
that most of the turbulent subgrid force $\bF_\ell$ comes, pointwise, from
pairs of velocity modes at length-scales $\sim \ell.$

The above arguments do not quite settle the issue of locality of the
circulation-flux $K_\ell(C,t),$ however. The delicate point here is that
large cancellations are expected in the line-integral of $\bF_\ell$ that
defines that flux. In order to infer scale-locality of $K_\ell(C,t),$ one
must assume that similar cancellations occur in the line integrals
of $\bF_\ell^{>\Delta}$ and $\bF_\ell^{<\delta}.$ This issue is hard to
address mathematically but may be investigated using data from
simulation or experiment.

The UV-locality properties of the subgrid force $\bF_\ell$ may be used
to develop an analytical expression for it, by means of a multi-scale
gradient expansion \cite{Eyink06a}. We consider only the lowest-order
term in that expansion, which corresponds to the so-called ``nonlinear
model'' for the stress \cite{MeneveauKatz00}:
\be \tau_{ij}=C\ell^2\Ol{u}_{i,l}\Ol{u}_{j,l}. \lb{nonlinear} \ee
Here $C=\int d\br |G(\br)|^2 r_1^2$ and a spherically-symmetric filter
function is assumed, so that $r_1$ could be replaced with any other
single component $r_i.$ (In terms of the constant $C_2$ employed in
\cite{Eyink06a}, $C=C_2/d$ where $d$ is the space dimension.) We use
the convention of subscript ``$,j$'' to denote $\partial_j,$ so that,
for example, $u_{i,j}=\partial u_i/\partial x_j.$ We also employ the
Einstein summation convention for repeated indices. To avoid an excess
of subscripts, we drop above and hereafter the subscript $\ell,$ since
a fixed filter length will be always understood. The physical assumption
behind the formula (\ref{nonlinear}) is strong UV-locality, so that
only adjacent subscale modes contribute to the stress. We expect that
this extreme assumption is fairly good in the $d=3$ energy cascade and
the $d=2$ direct enstrophy cascade. However, we present arguments below
that it fails badly for the $d=2$ inverse energy cascade. Note that it
is already known that the energy transfer is only weakly scale-local
in  $d=2$ \cite{Kraichnan71,Eyink06b,Chenetal06a}.

{}From the formula (\ref{nonlinear}) for the stress, one obtains the
corresponding formula for the subscale turbulent force:
\be f_i = -\partial_j\left(C\ell^2\Ol{u}_{i,l}\Ol{u}_{j,l}\right).
    \lb{nonlin-f1} \ee
By means of a standard vector calculus identity, this can be written
for $d\leq 3$ as:
\be f_i    = C\ell^2\epsilon_{ijk}\Ol{u}_{j,l}\Ol{\omega}_{k,l}
 -\partial_i\left(\frac{1}{2}C\ell^2\Ol{u}_{j,l}\Ol{u}_{j,l}\right).
 \lb{nonlin-f2} \ee
Here $\epsilon_{ijk}$ is the anti-symmetric Levi-Civita tensor for $d=3.$
This formula can be simplified by substituting $\Ol{u}_{j,l}=\Ol{S}_{jl}
-(1/2)\epsilon_{jlm}\Ol{\omega}_m$ in the first term and
$\Ol{u}_{j,l}\Ol{u}_{j,l}
=\Ol{S}_{j,l}\Ol{S}_{j,l} +\frac{1}{2}|\Ol{\bomega}|^2$ in the second,
yielding:
\be f_i =  C\ell^2\epsilon_{ijk}\Ol{S}_{jl}\Ol{\omega}_{k,l}
   -\partial_i\left(\frac{1}{2}C\ell^2\Ol{S}_{jl}\Ol{S}_{jl}\right).
\lb{nonlin-f3} \ee
This is our final formula for the turbulent force.
Substituting (\ref{nonlin-f3}) into (\ref{circ-flux}) yields a similar
formula for the circulation-flux:
\be K_\ell(C,t) = -C\ell^2 \oint_{\Ol{C}_\ell(t)}
          \epsilon_{ijk}\Ol{S}_{jl}\Ol{\omega}_{k,l}\,dx_i .
\lb{nonlin-circ-flux} \ee
According to this formula, the diffusion of vortex-lines out of the loop
is driven by strain acting upon the gradient of the vorticity vector.
This is plausible, since the turbulent force should act to smooth out
inhomogeneities in the large-scale vorticity field and become
negligible when the latter is constant.

The same result (\ref{nonlin-circ-flux}) for the circulation-flux can be
obtained from the ``nonlinear model'' of the turbulent vortex-force:
\begin{eqnarray}
 f_i^* & = & C\ell^2\epsilon_{ijk}\Ol{u}_{j,l}\Ol{\omega}_{k,l} \cr
       & = & C\ell^2\epsilon_{ijk}\Ol{S}_{jl}\Ol{\omega}_{k,l}
       + \partial_i\left(\frac{1}{4}C\ell^2|\Ol{\omega}|^2\right).
\lb{nonlin-fv} \end{eqnarray}
Although this derivation yields the same result, it is theoretically
less well-founded because of the poorer UV-locality properties
of the vortex-force. On the other hand, it gives a little more
physical insight, especially through the following alternative expression
for the vortex-force:
\be \bF^*= C\ell^2\grad\bdot(\Ol{\bS}\btimes\Ol{\bomega})
                 +\frac{1}{2}C\ell^2(\Ol{\bomega}\bdot\grad)\Ol{\bomega}.
\lb{nonlin-fv2}
\ee
Here
$(\Ol{\bS}\btimes\Ol{\bomega})_{ji}=\epsilon_{ikl}\Ol{S}_{jk}\Ol{\omega}_l$
defines what was termed in \cite{Eyink06a} the ``skew-strain matrix'' for
$d=3.$
Formula (\ref{nonlin-fv2}) is straightfowardly derived by calculating the
divergence $(\Ol{\bS}\btimes\Ol{\bomega})_{ji,j}$ and gathering the terms.
This expression makes a nice connection with the MSG expansion for the
turbulent stress, developed in \cite{Eyink06a}. The first term on the
righthand side of (\ref{nonlin-fv2}) corresponds to one of the stress
contributions in the MSG expansion, proportional to ``skew-strain''.
That term makes no strongly UV-local contribution to energy flux but a
major contribution to helicity flux and here we see also to circulation flux.

The second term on the righthand side of (\ref{nonlin-fv2})
corresponds to another term from the MSG expansion in \cite{Eyink06a},
a contractile stress along vortex-lines, $\tau^{{\rm vortex}}_{ij}\propto
-\Ol{\omega}_i\Ol{\omega}_j.$ As discussed in \cite{Eyink06a}, the effects
of the small-scale turbulence give the large-scale vortex-lines ``elastic''
properties. The second term in (\ref{nonlin-fv2}) therefore
has a simple geometric interpretation and can be written as
\begin{eqnarray}
\bF^{{\rm vortex}} &= &
       \frac{1}{2}C\ell^2(\Ol{\bomega}\bdot\grad)\Ol{\bomega} \cr
   & = & \frac{\partial}{\partial s}\left(\frac{1}{4}C\ell^2
|\Ol{\bomega}|^2\right)\widehat{\bomega}
+\frac{1}{2}\kappa C\ell^2|\Ol{\bomega}|^2\widehat{\bn}. \lb{nonlin-FS}
\end{eqnarray}
To derive (\ref{nonlin-FS}) we have used the Frenet-Serret equations
(e.g. see \cite{Struik61}) with $\widehat{{\bf t}}=\widehat{\bomega}$ the
unit tangent vector along large-scale vortex lines, $\widehat{{\bf n}}$
the unit normal vector and $\widehat{{\bf b}}$ the binormal \footnote{
Our notations here differ from those employed in section \ref{balance},
where $\widehat{{\bf t}}_\ell$ referred to the unit tangent vector to the loop
in the large-scale circulation and ${\bf n}_\ell=\widehat{{\bf t}}_\ell
\btimes\widehat{\bomega}_\ell.$}. The term $\bF_{\|}^{{\rm vortex}}$
in (\ref{nonlin-FS}) parallel to vortex lines arises from variations
in the vortex-strength along the line. The term $\bF_{\perp}^{{\rm vortex}},$
which arises from bending of vortex lines, is proportional to the curvature
$\kappa$ of the line and is directed along the normal $\widehat{{\bf n}}$.
Note that (\ref{nonlin-FS}) gives a contribution  to vorticity transport,
$\bsigma^{{\rm vortex}}=(1/2)\kappa C\ell^2|\Ol{\bomega}|^2\widehat{{\bf b}},$
which is directed along the binormal, reminiscent of the velocity of a
slender vortex filament in the local-induction approximation
\cite{Saffman1992}.

The formulas (\ref{nonlin-f3}) and (\ref{nonlin-fv}) for the turbulent
force simplify in space dimension $d=2.$ In that case,
\begin{eqnarray}
f_i & = & -C\ell^2\widetilde{\Ol{S}}_{ij}(\partial_j\Ol{\omega})
   -\partial_i\left(C\ell^2\Ol{\sigma}^2\right), \cr
f_i^* & = & -C\ell^2\widetilde{\Ol{S}}_{ij}(\partial_j\Ol{\omega})
      + \partial_i\left(\frac{1}{4}C\ell^2|\Ol{\omega}|^2\right),
\lb{nonlin-f-2D}
\end{eqnarray}
where $\pm \Ol{\sigma}$ are the eigenvalues of the symmetric,
traceless strain matrix $\Ol{S}_{ij}$ and $\widetilde{\Ol{S}}_{ij}=
\Ol{S}_{ik}\epsilon_{kj}=-\epsilon_{ik}\Ol{S}_{kj}$ is another symmetric,
traceless matrix, called in \cite{Eyink06b} the ``skew-strain matrix'' for
$d=2$. (Note that $\epsilon_{ij}$ is the $d=2$ anti-symmetric Levi-Civita
tensor.) The corresponding result for the circulation-flux is
\be K_\ell(C,t) = C\ell^2 \oint_{\Ol{C}_\ell(t)}
          \widetilde{\Ol{S}}_{ij}(\partial_j\Ol{\omega})\,dx_i .
\lb{nonlin-circ-flux-2D} \ee
This result can be derived as well from equation (\ref{circ-flux-dec})
and the ``nonlinear model'' for the vorticity transport vector in
$d=2,$
\be \sigma_i = C\ell^2 \Ol{u}_{i,j}(\partial_j\Ol{\omega}),
\lb{nonlin-sigma} \ee
previously considered in \cite{Eyink00,Chenetal03}. (This formula
is equivalent to that for the vortex-force in eq.(\ref{nonlin-f-2D}).)
Note, however, that the formula (\ref{nonlin-sigma}) predicts
``down-gradient'' transport of vorticity whenever there is a positive
rate of vorticity-gradient stretching and this is expected in $d=2$
both for the direct enstrophy cascade \cite{Eyink00,Chenetal03} and
also the inverse energy cascade \cite{Eyink06b,Chenetal06a}.
``Down-gradient'' vorticity transport is qualitatively correct
in the enstrophy cascade and there (\ref{nonlin-circ-flux-2D})
may yield a good approximation. However, in the inverse energy
cascade the vorticity transport must be ``up-gradient'' or
``anti-diffusive.'' Therefore, (\ref{nonlin-circ-flux-2D}) is
not likely to be a good approximation in the inverse cascade
range. It must be corrected by higher-order terms in the
convergent MSG expansion, corresponding to smaller subgrid
scales or higher-order gradients.

\section{Conclusions}
\lb{conclusion}

The main purpose of this paper was to elaborate a physical theory
of ``circulation cascade'' in classical fluid turbulence. We have
attempted to explain the conceptual basis of the phenomenon,
its physical mechanisms, the scale-locality properties of the
cascade, and its relation to inertial-range vortex-stretching
and energy transfer. Clearly, there are many important issues
that call for further work. Chief among these is to determine
the validity of G. I. Taylor's proposed mechanism for
turbulent energy-dissipation, based on vortex line-stretching
\cite{Taylor17,TaylorGreen37,Taylor1938}. Even after seventy
years of research, basic elements of Taylor's proposal remain
open to question. In particular, the strong inertial-range
violations of the Kelvin Theorem---predicted in \cite{Eyink06}
and observed in \cite{Chenetal06}---cast some doubt on a
key piece of Taylor's theory. It is possible that circulation
conservation remains valid in some weaker sense, e.g. the
conditional-mean version of the ``martingale conjecture'' in
\cite{Eyink06}. Further research is necessary to see whether
any weaker form of the Kelvin Theorem holds at high Reynolds
numbers and, if so, whether it is sufficient for the purposes
of Taylor's  mechanism. It should be emphasized that even the
{\it existence} of circulations in the infinite-Reynolds-number
limit is an open question. Advected loops in a turbulent flow
are expected to become fractal \cite{Mandelbrot76,
SreenivasanMeneveau1986} and defining line-integrals for
non-rectifiable curves demands some mathematical sophistication
\cite{Eyink06}. In superfluids the advected contours in Kelvin's
Theorem can also become highly distorted, with interesting
consequences for vortex motion \cite{Nilsenetal05}. Fractality
of the advected loops could have significant implications
for conservation of circulations in fluid turbulence.

In addition to hydrodynamics of incompressible fluids, there
are other turbulent systems for which phenomena similar
to ``circulation cascade'' are expected to exist. Of these,
one of the most significant is magnetohydrodynamic (MHD) turbulence
of plasmas. In this case it is Alfv\'{e}n's Theorem \cite{Alfven43}
on conservation of magnetic flux in the ideal, zero-resistivity limit
which plays the role of Kelvin's Theorem. However, there is strong
evidence from observations of magnetic flux reconnection rates
in astrophysical settings to believe that Alfv\'{e}n's Theorem breaks
down in MHD turbulence even with negligible resistivity \cite{Dere96,
LazarianVishniac00}. This violation of conservation of magnetic
flux, presumably due to a similar cascade phenomenon as
for Navier-Stokes dynamics, is discussed in a following paper
for the MHD equations \cite{EyinkAluie06}. Another important
problem is turbulence in geophysical fluids, where Ertel's Theorem
\cite{Ertel42} on conservation of potential vorticity (PV) plays
a fundamental role in theories of quasi-geostrophy. It is
well-known that Ertel's Theorem is a differential form of the
Kelvin Theorem (e.g. see \cite{Pedlosky87}, Section 2.5 or
\cite{Salmon1988}, Section 4). The ``cascade of circulations''
in this context should be quite similar, generally speaking,
to that for two-dimensional Navier-Stokes and correspond to a
turbulent transport of PV out of the advected loop. However,
in geophysical fluid dynamics there is an additional complication
that the loop in Kelvin's theorem must lie in a surface of constant
density (bouyancy) or pressure \cite{Pedlosky87,Salmon1988}. Thus,
turbulent mixing of isopycnal surfaces is an additional source of breakdown
of Kelvin's Theorem in this context. Finally, another interesting
setting for ``circulation cascade'' is superfluid turbulence
\cite{Barenghietal01}. The analogy between quantum phase-slip
and circulation-cascade could prove useful here.

\begin{acknowledgments}
I wish to thank S. Chen, M. Wan and Z. Xiao
for a very fruitful collaboration on this problem and R. Ecke, N. Goldenfeld,
D. D. Holm, S. Kurien, C. Meneveau, S. Nazarenko, A. Newell, K. R. Sreenivasan,
A. Tsinober, E. Vishniac  and B. Wingate for conversations.  I am particularly
grateful
to P. Ao for suggesting a connection between Onsager's ``dissipative anomaly''
and the Josephson-Anderson relation in superfluids.  This work was supported
by NSF grant \# ASE-0428325 at the Johns Hopkins University and by the Center
for Nonlinear Studies at Los Alamos National Laboratory.
\end{acknowledgments}

\bibliography{circulation}% Produces the bibliography via BibTeX.

\end{document}